\documentclass[conference]{IEEEtran}
\usepackage[utf8]{inputenc}
\usepackage[english]{babel}
\usepackage{cite}
\usepackage{amsmath,amssymb,amsfonts}
\usepackage{algorithmic}
\usepackage{graphicx}
\usepackage{textcomp}
\usepackage{xcolor}
\usepackage{hyperref}
\usepackage{float}
\usepackage{listings}
\usepackage{booktabs} 
\usepackage{amssymb}  
\definecolor{codegreen}{rgb}{0,0.6,0}
\definecolor{codegray}{rgb}{0.5,0.5,0.5}
\definecolor{codepurple}{rgb}{0.58,0,0.82}
\definecolor{backcolour}{rgb}{0.95,0.95,0.92}

\lstdefinestyle{mystyle}{
    backgroundcolor=\color{backcolour},   
    commentstyle=\color{codegreen},
    keywordstyle=\color{magenta},
    numberstyle=\tiny\color{codegray},
    stringstyle=\color{codepurple},
    basicstyle=\ttfamily\footnotesize,
    breakatwhitespace=false,         
    breaklines=true,                 
    captionpos=b,                    
    keepspaces=true,                 
    numbers=left,                    
    numbersep=5pt,                  
    showspaces=false,                
    showstringspaces=false,
    showtabs=false,                  
    tabsize=2
}
\begin{document}

\title{High-Dimensional Data Processing: Benchmarking Machine Learning and Deep Learning Architectures in Local and Distributed Environments}

\author{
    \IEEEauthorblockN{
        José Julián Rodríguez Gutiérrez,
        Piotr Enriquevitch Lopez Chernyshov,
        Emiliano Lerma García,\\
        Rafael Medrano Vazquez, and
        Jacobo Hernández Varela
    }
    \IEEEauthorblockA{
        {\itshape Licenciatura en Ingeniería de Datos e Inteligencia Artificial}\\
        {\itshape División de Ingenierías Campus Irapuato-Salamanca}\\
        Guanajuato, México\\
        {\footnotesize\{jj.rodriguez.gutierrez, pe.lopezchernyshov, e.lermagarcia, r.medranovazquez, j.hernandezvarela\}@ugto.mx}
    }
}

\maketitle

\begin{abstract}
This document reports the sequence of practices and methodologies implemented during the Big Data course. It details the workflow beginning with the processing of the Epsilon dataset through group and individual strategies, followed by text analysis and classification with RestMex and movie feature analysis with IMDb. Finally, it describes the technical implementation of a distributed computing cluster with Apache Spark on Linux using Scala.
\end{abstract}


\section{Introduction}
The present study develops as an integral big data project encompassing three complementary analyses across different domains. The first case study, centered on the Epsilon dataset, served as an introduction to massive data handling through the implementation of a Multi-Layer Perceptron (MLP) with architecture 2000-128-128-2, trained on 100,000 instances with 2,000 features. Using PyTorch with GPU acceleration (CUDA), the model achieved 88.98\% accuracy after 100 training epochs, with a batch size of 128 and Adam optimization (learning rate=$10^{-5}$, weight decay=$10^{-4}$). The implementation included regularization techniques through Batch Normalization and Dropout ($p=0.8$), reducing validation loss from 0.6847 to 0.2670. This exercise established the methodological foundations for efficient processing of high-dimensionality datasets.

The second analysis, focused on the Rest-Mex dataset, implemented a complete pipeline for polarity classification in Mexican tourist reviews. Text preprocessing techniques were applied including tokenization, stopword removal, and lemmatization, followed by vectorization using CountVectorizer or TF-IDF. The supervised classification model categorized reviews into three classes (Positive/Negative/Neutral), addressing the inherent imbalance of the tourist dataset through class weighting techniques.

Finally, the IMDb dataset analysis represents the project's culmination, integrating deep textual analysis through TF-IDF (5,000 features with HashingTF and minDocFreq=3) over descriptions, titles, and metadata of 85,855 films. An intelligent contextual imputation system was implemented based on director, genre, actor, and writer averages, significantly reducing missing values in critical variables. Optimized XGBoost Regressor through Grid Search (36 hyperparameter combinations) with 3-fold cross-validation achieved an RMSE of 0.6001 and $R^2$ of 0.79 in continuous rating prediction. The optimal hyperparameters found were: maxDepth=12, eta=0.03, numRound=500, minChildWeight=3.

Sentiment analysis of descriptions revealed a distribution of 46.83\% neutral, 32.65\% positive, and 20.52\% negative, information that was incorporated as an additional feature. Exploratory analysis identified Christopher Nolan (8.22), Satyajit Ray (8.02), and Hayao Miyazaki (8.01) as the directors with the highest average ratings, while Film-Noir (6.64), Biography (6.62), and History (6.54) emerged as the best-rated genres.

This methodological progression—from binary classification in Epsilon, through sentiment analysis in Rest-Mex, to continuous rating prediction in IMDb—demonstrates the versatility of machine learning techniques applied to different scales and data types, establishing a robust framework for big data analysis across multiple domains.

\section{State of the Art}

\subsection{Stochastic Gradient Descent for Large-Scale SVM(Epsilon)}

In the field of large-scale binary classification, Shalev-Shwartz et al. (2011) developed Pegasos, a stochastic sub-gradient descent algorithm for SVM that proved especially efficient for massive datasets. The algorithm achieves a complexity of $\tilde{O}(1/\lambda\epsilon)$ iterations to reach $\epsilon$ accuracy, where each iteration operates on a single training example. This efficiency makes it particularly suitable for large-scale text classification problems.

Pegasos' approach works directly on the primal objective, avoiding the need to maintain the complete kernel matrix in memory, which represents a significant advantage when working with high-dimensionality datasets like Epsilon. Reported experiments show speedups of up to an order of magnitude over conventional SVM methods on datasets with sparse features, achieving accuracies above 85\% in binary classification problems with millions of examples. The work establishes that the algorithm's convergence does not directly depend on the training set size, but rather on the regularization parameter $\lambda$, making it scalable for big data applications.

\subsection{Movie Success Prediction in Bollywood(IMDB)}

Verma and Verma (2019) developed predictive models for Bollywood films using supervised algorithms such as Random Forest, Logistic Regression, SVM, and Adaptive Tree Boosting, achieving 92\% accuracy in binary classification (success/failure). Their work identified music rating, IMDb ranking, and number of release screens as key predictors, working with a dataset of 200 films and employing cross-validation to evaluate model performance.

\subsection{Sentiment Analysis in Mexican Tourism Texts(Rest-Mex)}

In the Latin American context, Álvarez-Carmona et al. (2023) developed the Rest-Mex task at IberLEF 2023, focusing on sentiment analysis of Mexican tourist texts extracted from TripAdvisor. The dataset comprised 359,565 reviews classified by three simultaneous attributes: ordinal polarity (1-5 scale, where 1=Very bad, 5=Very good), attraction type (hotel, restaurant, tourist attraction), and visited country (Mexico, Cuba, Colombia). The corpus presented marked class imbalance, with 157,095 reviews of polarity 5 compared to only 5,772 of polarity 1 in the training set.

The best performing systems, based on Spanish pre-trained RoBERTa with tourism domain adaptation, achieved a weighted F-measure of 0.78 for polarity, 0.99 for attraction type, and 0.94 for country classification. The LKE-IIMAS team implemented data augmentation techniques through back translation to balance minority classes, demonstrating the effectiveness of contextual transformers in multi-class classification tasks with Spanish textual data. The study also proposed a novel "easiness" metric based on set operations, finding that polarity was the most difficult attribute to predict (easiness=0.38) while attraction type was the easiest (easiness=0.50).

\section{Epsilon Dataset Description}

The Epsilon Dataset stands out as a high-performance benchmark dataset, designed to challenge algorithms in large-scale, high-dimensional \textbf{binary classification} tasks.

\subsection{Technical Characteristics}
The dataset comprises 400,000 samples for training and 100,000 for validation. Each record is described by 2001 columns: the first serves as the class label ($\{0, 1\}$) and the remaining 2000 represent \textbf{numerical} features with real (floating-point) values. The total volume of the complete dataset is estimated at \textbf{11 GB}. For the practices in this course, the dataset was segmented into five equal parts to facilitate resource management and evaluation.

\subsection{Purpose as Benchmark}
Epsilon was generated through \textbf{complex and anonymized transformations} of original data (presumably in risk modeling or financial domains). Its objective is not the semantics of the features, but rather the computational challenge it presents:
\begin{itemize}
    \item \textbf{Scalability:} Evaluate the efficiency of \textit{machine learning} algorithms and \textbf{distributed computing} systems (such as Apache Spark) under massive and dense data loads.
    \item \textbf{High Dimensionality:} Serves as a rigorous test for \textbf{regularization} and overfitting management in models with a large number of input variables (2000), where correlations are subtle.
\end{itemize}

\section{Epsilon Dataset Processing: Individual Approach}

Individual processing of Epsilon allowed for the evaluation of various machine learning algorithm performance in a local environment, leveraging frameworks such as PyTorch \cite{paszke2019pytorch} and scikit-learn to obtain detailed and comparable results.

\subsection{Model Selection Strategy}

The selection of the best model was based on a combined evaluation strategy to mitigate the impact of potentially biased splits in the dataset. The dataset was divided into 5 parts, and a specific combination of two algorithms was assigned to 5 "Persons" or evaluation instances (A, B, C, D, E), seeking a more robust view of each algorithm's behavior.

The following table summarizes the work division and algorithm combinations used in each instance:

\begin{table}[h!]
    \centering
    \caption{Algorithm Assignment by Evaluation Instance}
    \begin{tabular}{ll}
        \toprule
        Model & Algorithms Evaluated \\
        \midrule
        Person A & Logistic Regression + Random Forest \\
        Person B & MLP + Logistic Regression \\
        Person C & XGBoost + MLP \\
        Person D & SVM + XGBoost \\
        Person E & SVM + Random Forest \\
        \bottomrule
    \end{tabular}
    \label{fig:repartition_algorithms}
\end{table}

The fundamental reason for this division is to obtain a more robust view of each algorithm's behavior. By testing algorithms on different instances of the dataset (the 5 divided parts), we seek to determine whether a model's performance is consistently good or strongly influenced by a specific partition. If an algorithm performs well despite variations inherent to dataset divisions, it provides strong evidence that it will be the best-performing model on the complete dataset.

\subsection{Architecture and Results of the Winning Model (MLP)}

After evaluation, the model that demonstrated the best average performance was the Multi-Layer Perceptron (MLP) \cite{goodfellow2016deep}. The MLP architecture, previously defined in PyTorch code, consists of a sequence of layers including a Linear Layer (FC), Batch Normalization (BN), Activation Function (ReLU), and Regularization (Dropout) in each hidden block.

\begin{figure}[H]
    \centering
    \includegraphics[width=0.4\columnwidth]{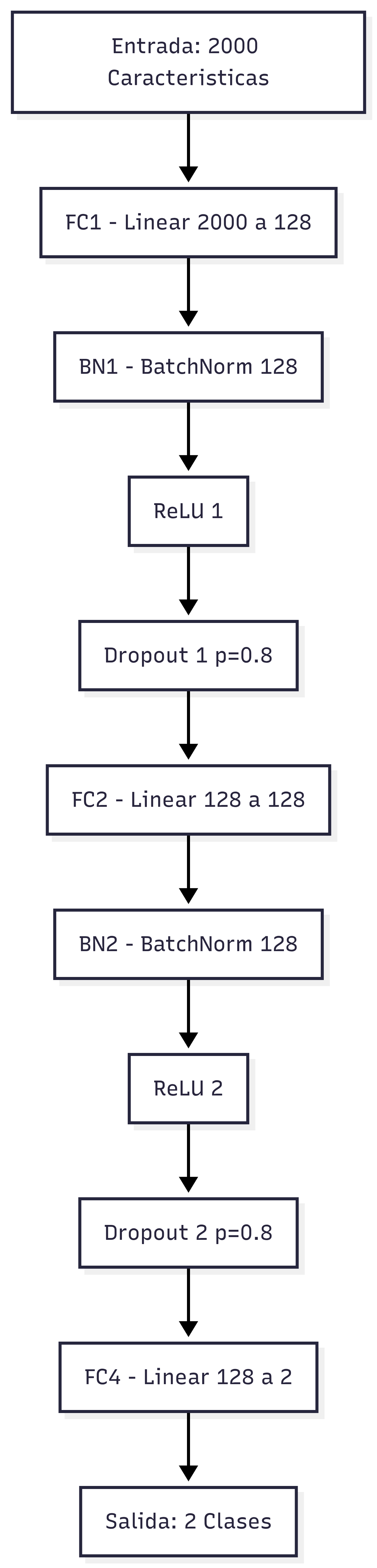}
    \caption{MLP architecture flow.}
    \label{fig:mlp_arch}
\end{figure}

\paragraph{The training hyperparameters used for the MLP were:}

\begin{itemize}
    \item $input\_size = 2000$ (number of features).
    \item $hidden\_size = 128$ (this parameter is inferred from $input\_size$).
    \item $output\_size = 2$ (for binary classification).
    \item $learning\_rate = 1 \times 10^{-5}$ (learning rate).
    \item $weight\_decay = 1 \times 10^{-4}$ (L2 regularization).
    \item $epochs = 100$ (number of training epochs).
    \item $batch\_size = 128$ (batch size).
    \item Loss Function: $nn.CrossEntropyLoss()$.
    \item Optimizer: $optim.Adam$ (with $lr = \text{learning\_rate}$ and $\text{weight\_decay}=\text{weight\_decay}$).
\end{itemize}

\subsection{Individual Evaluation Results with K-Fold Cross-Validation}

The final evaluation of all models was performed using K-Fold Cross-Validation \cite{kohavi1995study} with $K=5$ folds, averaging the performance metrics.

\begin{table}[H]
    \centering
    \caption{Average Model Results with 5-Fold Cross-Validation (Individual Processing)}
    \resizebox{\columnwidth}{!} {
    \begin{tabular}{lcccc}
        \toprule
        Model & Average Accuracy (\%) & Macro F1-Score (\%) & AUC-ROC (\%) & Training Time (s) \\
        \midrule
        MLP (Baseline) & 89.18 & 89.18 & \textbf{93.00} & 503.23 \\
        XGBoost & 88.55 & 88.53 & 91.80 & 847.91 \\
        Random Forest & 87.90 & 87.87 & 90.50 & 782.45 \\
        SVM & 85.12 & 85.08 & 88.20 & 1054.67 \\
        Logistic Regression & 84.50 & 84.49 & 87.50 & 487.58 \\
        \bottomrule
    \end{tabular}
    }
\end{table}

\begin{figure}[H]
    \centering
    \includegraphics[width=0.45\columnwidth]{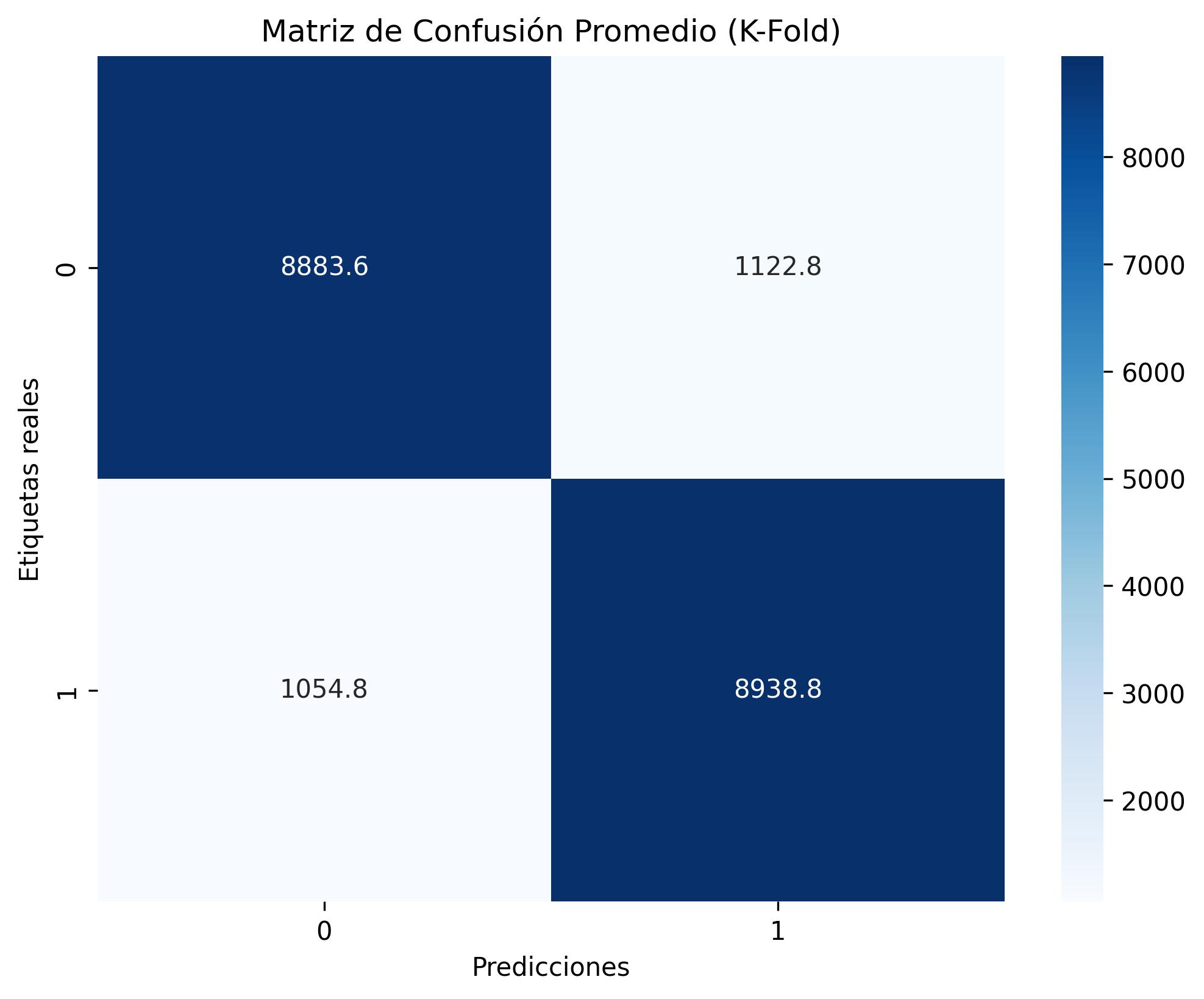}
    \caption{Average Confusion Matrix of MLP (Individual Processing).}
    \label{fig:conf_matrix_individual}
\end{figure}

\begin{figure}[H]
    \centering
    \includegraphics[width=0.45\columnwidth]{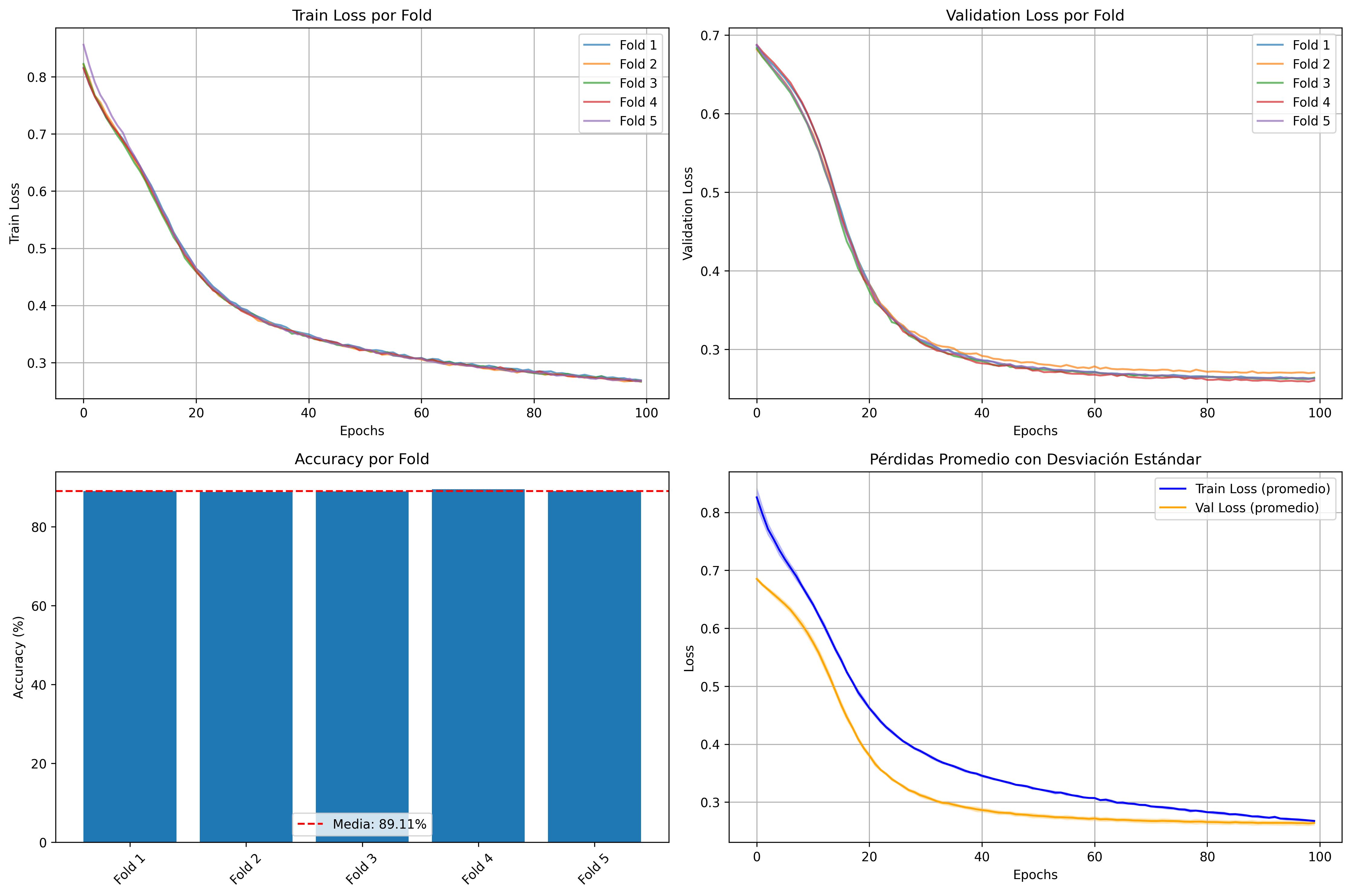}
    \caption{MLP Learning Curve (Individual Processing).}
    \label{fig:learning_curve_individual}
\end{figure}

\subsection{Analysis of Individual Processing Results}

The analysis of the results table indicates that the MLP not only has the best average Accuracy (89.18\%), but also demonstrates the greatest consistency in Macro Recall and Macro F1-Score, suggesting an optimal balance between prediction precision and the ability to correctly identify positive instances of both classes.

The Confusion Matrix (Figure \ref{fig:conf_matrix_individual}) shows that the model has almost symmetric performance in classifying both classes, which validates the high Macro F1-Score, but this gives us an indication that the data that has been misclassified is very similar and generates this confusion. For better performance, it would be necessary to analyze this data to find out what makes them so similar.

Finally, the Learning Curve (Figure \ref{fig:learning_curve_individual}) shows that the model is indeed very robust since both the train and test curves do not present data memorization and the curves converge constantly to a certain point. The models of all 5 Folds are consistent with each other, there is no one with strange behavior for this block, which gives us high reliability in our best model.

\section{Implementation of Distributed Infrastructure: Apache Spark Cluster}

\subsection{Motivation and Focus}

To overcome the computational limitations (RAM memory and processing capacity) encountered during local (single-node) analysis of the voluminous Epsilon dataset \cite{chang2011libsvm}, a distributed processing architecture was designed and implemented. The objective was not only to scale processing, but also to comparatively evaluate the performance gains and feasibility of using a home cluster for large-scale data preparation and analysis tasks.

\subsection{Hardware Architecture and Configuration}

A cluster based on Apache Spark's Master-Worker architecture \cite{zaharia2010spark} was deployed, consisting of four nodes interconnected in a dedicated local area network (LAN). This topology allowed the workload of the Epsilon dataset to be distributed across multiple machines.

\begin{table}[H]
    \centering
    \caption{Hardware specifications of the Apache Spark cluster implemented.}
    \label{tab:cluster-hardware-specs}
    \begin{tabular}{llllll}
        \toprule
        Node Role & Quantity & CPU (Cores) & RAM  & Storage & OS \\ \midrule
        Master       & 1        & 2           & 16GB & 128GB          & Ubuntu  \\
        Workers      & 3        & 8           & 16GB & 512GB          & Ubuntu  \\ \bottomrule
    \end{tabular}
\end{table}

Communication between nodes was configured using SSH (Secure Shell) without a password (password-less SSH) to enable automated management of Spark daemons from the master node. Static IP addresses were assigned to each node to ensure network topology stability.

\subsection{Software Configuration}

The software environment was standardized across all nodes to avoid version conflicts and serialization errors. Apache Spark 3.5.7 was used on JVM (Java Virtual Machine) v21.0.9.

The cluster configuration was performed in Standalone mode. Critical parameters adjusted in the spark-env.sh and spark-defaults.conf files to optimize the use of specific resources of the available hardware included:

\begin{itemize}
    \item SPARK\_WORKER\_MEMORY: It was limited to 8 GB per worker node, reserving memory for the operating system and other background processes.
    \item SPARK\_WORKER\_CORES: The 8 physical cores available in each worker were assigned to maximize thread-level parallelism.
    \item spark.driver.memory: It was increased to 4 GB on the master node to efficiently handle the aggregation of intermediate results and prevent memory overflows in the reduce phase.
\end{itemize}

\subsection{Pipeline Implementation and Execution}

The processing and analysis pipeline, initially designed for a local environment, was adapted for distributed execution. A key decision was data management: to minimize network latency and leverage the principle of data locality, a complete copy of the Epsilon dataset was preloaded into the local storage of each worker node.

The execution was orchestrated from the master node using the spark-submit tool. The application code, written in Scala (including its dependencies), was packaged in a JAR file. The command explicitly pointed to the cluster manager (spark://[MASTER-IP]:7077), thus distributing the execution of transformations (such as filter and vectorization) and actions (count, train model) on the Epsilon dataset among the three worker nodes.

\subsection{Performance Results and Analysis}

The final validation of the distributed infrastructure focused on a key objective: training models on the entire Epsilon dataset without resorting to sampling or forced partitioning. While the local environment was limited by physical memory, forcing batch processing, the Spark cluster was able to ingest and process the entire dataset in a single unified run, demonstrating its ability to scale horizontally.

The following are the performance metrics obtained when running four classification algorithms on the distributed cluster over the entire dataset:

\begin{table}[H]
    \centering
    \caption{Distributed Cluster Training Metrics}
    \label{tab:cluster-results}
    \begin{tabular}{lcc} \toprule
        \textbf{Algorithm} & \textbf{Training Time (s)} & \textbf{AUC-ROC} \\ \midrule
        Linear SVC & 136.57 & 0.9402 \\
        Logistic Regression & 166.78 & \textbf{0.9504} \\
        Random Forest & 181.52 & 0.7778 \\
        Gradient Boosted Trees & 203.79 & 0.8075 \\ \bottomrule
    \end{tabular}
\end{table}

\subsubsection{Comparative Analysis of Results}

Data analysis reveals important trends regarding algorithm performance in a distributed environment with large-scale data:

\begin{itemize}
    \item \textbf{Computational Efficiency:} Linear SVC is the most efficient algorithm, completing training in 136.57 seconds. This result highlights the cluster's ability to optimally parallelize intensive linear algebra and convex optimization operations specific to this model \cite{meng2016mllib}.
    
    \item \textbf{Balance between Speed and Accuracy:} Although SVC was the fastest, Logistic Regression offered the best balance between speed and predictive quality. With a time penalty of approximately 30 additional seconds, it achieved the highest AUC-ROC (0.9504), positioning it as the most robust option for this dataset in its base configuration.

    \item \textbf{Behavior of Tree-Based Algorithms:} The Random Forest and Gradient Boosted Trees (GBT) algorithms showed significantly higher computational cost (exceeding 180 seconds) and inferior predictive performance (AUC-ROC < 0.81). This phenomenon suggests that, for the dimensionality and nature of the Epsilon dataset, linear models scale and generalize better in a distributed configuration without exhaustive hyperparameter tuning. The complexity of the trees, combined with high dimensionality, may have led to overfitting \cite{breiman2001random} or required more iterations to converge.
\end{itemize}

\subsubsection{Conclusion of the Distributed Experiment}

The Apache Spark cluster demonstrated robustness and scalability by successfully completing tasks that were unfeasible in a single-node local environment. The distributed architecture fundamentally transformed the processing capabilities by maintaining the entire dataset in distributed memory across workers, effectively leveraging the aggregate RAM capacity of the cluster. This distributed memory model completely eliminated the Java Heap Space errors that had severely constrained local processing, where memory limitations forced data partitioning and sequential batch processing. Furthermore, the cluster infrastructure enabled a continuous and stable machine learning workflow on large-scale data, eliminating the interruptions and manual interventions required in the local environment, thus laying the foundation for more agile and iterative experimentation cycles. This experiment validates not only the technical feasibility of distributed computing for datasets of Epsilon's scale, but also demonstrates the practical advantages in terms of reliability, efficiency, and scalability that a distributed paradigm offers when analyzing datasets of such volume and complexity.

\section{Discusion Individual vs Distributed}

The comparative analysis confirms critical findings across both processing paradigms:

\textbf{Individual Processing:} The MLP emerged as the best-performing model with an average Accuracy of 89.18\% and AUC-ROC of 93.00\%, demonstrating superior predictive capability while maintaining competitive training time (503.23s). This optimal balance between accuracy and computational efficiency validates the effectiveness of deep learning approaches for high-dimensional classification tasks. Tree-based models (XGBoost with 91.80\% AUC-ROC and Random Forest with 90.50\% AUC-ROC) showed robust performance but required 68\% and 55\% more training time respectively without surpassing the neural network's accuracy. Linear models presented contrasting characteristics: Logistic Regression achieved the fastest training (487.58s) but with the lowest AUC-ROC (87.50\%), while SVM demonstrated the worst efficiency ratio with the longest training time (1054.67s) and only 88.20\% AUC-ROC, highlighting the computational challenges of kernel-based methods with large-scale, high-dimensional data.

\textbf{Distributed Processing:} The Apache Spark cluster successfully overcame local hardware limitations, enabling complete dataset processing without sampling or memory constraints. In the distributed environment, Logistic Regression achieved the best balance with an AUC-ROC of 95.04\% in just 166.78 seconds, representing a significant improvement over its local counterpart (from 87.50\% to 95.04\% AUC-ROC). Linear SVC demonstrated exceptional computational efficiency (136.57s) with strong predictive performance (94.02\% AUC-ROC), benefiting substantially from Spark's distributed linear algebra optimizations. This validates the practical advantage of distributed computing for large-scale data analysis, particularly for algorithms that can effectively leverage parallelization.

The performance gap between local and distributed implementations reveals an important insight: linear models in Spark MLlib achieved higher AUC-ROC scores (95.04\% and 94.02\%) compared to the local MLP (93.00\%), suggesting that access to the complete dataset without forced partitioning significantly benefits model generalization, potentially offsetting the representational advantages of neural networks for this particular dataset.

\section{RestMex: Polarity Classification}

\subsection{Problem Context}

As part of the exploration and learning process within the Big Data field, the task for the \textbf{Rest Mex 2023} challenge was undertaken. This challenge consists of a dataset containing reviews of establishments located in areas designated as ``Pueblos Mágicos'' (Magical Towns); the platform from which these opinions were extracted was TripAdvisor.

The objective was to classify opinions within a polarity spectrum according to a 5-star metric. This metric is shown below, along with the number of data instances per class (a volume that presents a challenge for correct processing).

\begin{table}[H]
    \centering
    \caption{Classification Metric}
    \label{tab:polaridad}
    \begin{tabular}{lc} 
        \toprule
        \textbf{Polarity} & \textbf{Stars} \\
        \midrule
        Very Bad   & $\bigstar$ \\
        Bad        & $\bigstar\bigstar$ \\
        Neutral    & $\bigstar\bigstar\bigstar$ \\
        Good       & $\bigstar\bigstar\bigstar\bigstar$ \\
        Very Good  & $\bigstar\bigstar\bigstar\bigstar\bigstar$ \\
        \bottomrule
    \end{tabular}
\end{table}

\begin{table}[H]
    \centering
    \caption{Instance Distribution by Class (Polarity)}
    \label{tab:distribucion_polaridad}
    \begin{tabular}{lr} 
        \toprule
        \textbf{Class} & \textbf{Instances} \\
        \midrule
        Very Bad & 5,441 \\
        Bad & 5,496 \\
        Neutral & 15,519 \\
        Good & 45,034 \\
        Very Good & 136,561 \\
        \midrule
        \textbf{Total} & \textbf{208,051} \\
        \bottomrule
    \end{tabular}
\end{table}

\subsection{Challenge and Techniques Used}
As observed in the technical datasheet of the presented dataset, there are two main components that make this task a suitable challenge for Big Data topics and even NLP (Natural Language Processing):

\begin{itemize}
    \item \textbf{Data type and volume:} The quantity proves to be a processing issue; additionally, the textual data requires preprocessing to be utilized effectively.
    
    \item \textbf{Class Imbalance:} There is a clear class imbalance regarding the "Very Good" (5 stars) instances, which can strongly bias any applied model.
\end{itemize}

\subsection{Preprocessing}
Before creating a classification model, a data exploration was conducted beyond what has been presented so far, identifying positive and negative ratings with duplicate attached reviews or those posted more than once. Although these could be valid data points, it was decided to treat them as ``SPAM'' and remove this data, as well as extremely short or incoherent opinions that could represent noise.

Even with this measure, the imbalance remained quite significant, so it was decided to apply data generation and reduction strategies: augmenting the minority class and reducing the majority class.

\subsection{Data Balancing Strategies}

\subsubsection*{Data Downsampling: Ring Technique}
To manage the majority classes (primarily labels 4 and 5), the \textit{Ring Technique} was applied. This undersampling method allows for reducing the data volume while preserving the spatial structure of the distribution. Unlike random deletion, this technique selects representative instances in different "rings" or density levels, ensuring that the intrinsic variability of the class is maintained despite the reduction in samples.

\subsubsection*{Data Augmentation: Backtranslation}
For the minority classes, the \textit{Backtranslation} data augmentation technique was used. This process consisted of translating the original Spanish sentences into a pivot language (English) and subsequently re-translating them back to Spanish. This strategy generated synthetic paraphrases that introduce lexical and grammatical diversity, enriching the training process without altering the semantics or the original polarity of the text.

\subsection{Experimental Results - REST MEX 2023}

For system validation, fine-tuning of the pre-trained model \textbf{BETO} was performed for 2 epochs, using a batch size of 16 and a learning rate of $1\times10^{-5}$. The total training time was 1.6 hours.

Table \ref{tab:resultados_beto} presents the model's performance on the test sets.

\begin{table}[H]
    \centering
    \caption{Evaluation Metrics (BETO Model)}
    \label{tab:resultados_beto}
    \begin{tabular}{lcc}
        \toprule
        \textbf{Task} & \textbf{Accuracy} & \textbf{F1-Score} \\
        \midrule
        Polarity & 86.26\% & 53.64\% \\
        \bottomrule
    \end{tabular}
\end{table}

\section{Processing and Classification: IMDb Dataset}
The IMDb dataset was analyzed with the objective of predicting movie ratings based on their metadata. It is important to note that this section was carried out using Spark and the Scala programming language.

\subsection{Dataset}
The original dataset \texttt{IMDb\_movies.csv} contains 85,855 records and 22 variables, including:
\begin{itemize}
    \item Textual variables: title, actors, director, writer, description.
    \item Numerical and categorical variables: duration, genre, votes, year.
    \item Revenue-related variables: budget, domestic income, and worldwide income.
\end{itemize}

\subsection{Feature Selection}
To classify the rating according to movie characteristics, the following variables were selected:
\begin{itemize}
    \item Genre
    \item Budget
    \item Main actors
    \item Production year
    \item Duration
    \item Votes
    \item Director
    \item Writer
    \item Production company
    \item Description
\end{itemize}

These features were selected through multiple prediction tests to determine the contribution of each one.

\subsection{EDAs}
The results obtained during the EDA (TOP 3) were:

\begin{itemize}
    \item \textbf{Top directors based on average vote}
    \begin{itemize}
        \item Christopher Nolan (8.22)
        \item Satyajit Ray (8.02)
        \item Hayao Miyazaki (8.01)
    \end{itemize}

    \item \textbf{Lowest-rated directors}
    \begin{itemize}
        \item Brett Kelly (1.53)
        \item Rene Perez (2.81)
        \item Jared Cohn (2.94)
    \end{itemize}

    \item \textbf{Highest-rated genres}
    \begin{itemize}
        \item Film-Noir (6.64)
        \item Biography (6.62)
        \item History (6.54)
    \end{itemize}

    \item \textbf{Lowest-rated genres}
    \begin{itemize}
        \item Horror (4.83)
        \item Sci-Fi (5.07)
        \item Thriller (5.47)
    \end{itemize}

    \item \textbf{Unique directors:} 34,733

    \item \textbf{Unique writers:} 66,859

    \item \textbf{Sentiment analysis on descriptions}
    \begin{itemize}
        \item 46.83\% Neutral
        \item 32.65\% Positive
        \item 20.52\% Negative
    \end{itemize}
\end{itemize}

\subsection{Cleaning of Numerical Features}
The numerical columns were cleaned by removing symbols such as \texttt{\$} and \texttt{,}. Afterwards, they were converted to float type to avoid datatype issues.  
Additionally, the variable \texttt{year} was normalized.

\subsection{Intelligent Imputation Through Contextual Mean}
After the EDA process, we observed that the columns \texttt{REVIEWS\_FROM\_USERS} and \texttt{REVIEWS\_FROM\_CRITICS} contained a large number of missing values: 7,597 and 11,797 respectively.  
Because reviews represent an important part of our features, we implemented a contextual imputation algorithm based on actors, genre, writer, and director to fill in these values.

\subsection{Text Processing}
A unified field named \texttt{all\_text} was generated by concatenating: title, genre, director, writer, production company, actors, and description.  
TF–IDF was then applied to this column. This process included Tokenization, Stopword Removal, HashingTF with 5,000 features, and IDF with \texttt{minDocFreq = 3}.

\subsection{Integration of Numerical Columns}
After text processing, a \texttt{VectorAssembler} generated the final feature vector using both TF–IDF text features and numerical features such as year, duration, and votes.  
This allowed the prediction model to have more information and therefore perform more efficiently.

\subsection{Model Selection}
After multiple prediction tests, we observed that the algorithm with best performance and efficiency was the \textit{XGBoost Regressor}.  
For this reason, it was selected as the final prediction algorithm. A Grid Search was implemented with 36 hyperparameter combinations varying depth, \texttt{numRound}, \texttt{eta}, and \texttt{minChildWeight}.  
Validation used 3-fold cross-validation.  
After this process, the optimal parameters found in our case were:

\begin{itemize}
    \item Gamma = 0.1 (Reduces unnecessary splits)
    \item maxDepth = 10 (Maximum tree complexity)
    \item eta = 0.05 (Learning rate)
    \item numRound = 300 (Number of boosting rounds)
    \item Lambda = 1.5 (L2 regularization for stability)
    \item minChildWeight = 5 (Minimum samples per leaf)
\end{itemize}

\subsection{Model Results}
\begin{itemize}
    \item RMSE = 0.6001
    \item MAE = 0.4927
    \item R2 = 0.7901
\end{itemize}

We observed that the model improves significantly compared to the simple non-optimized version, increasing the R2 by more than 10\%.


\section{Conclusions}

The chronological evolution of our methodology—from binary classification with Epsilon, through multilingual sentiment analysis with Rest-Mex, to continuous rating prediction with IMDb—represents more than a simple technical progression; it embodies the natural learning curve that data scientists and engineers must navigate when confronting real-world big data challenges. This pedagogical approach proved essential in understanding not only the tools and algorithms, but also the fundamental principles of scalability, distributed computing, and the trade-offs between model complexity and computational efficiency.

The Epsilon dataset analysis established the foundational infrastructure for big data processing, revealing critical insights about algorithm scalability. The Multi-Layer Perceptron achieved 89.18\% accuracy locally, while the distributed Apache Spark implementation enabled Logistic Regression to reach 95.04\% AUC-ROC by processing the complete dataset without sampling constraints. This 7.54 percentage point improvement demonstrates a key principle in big data analytics: access to complete, unsampled datasets can fundamentally transform model performance, sometimes offsetting the representational advantages of more complex architectures. The successful deployment of a physical Spark cluster using Scala eliminated the memory constraints that had limited local processing, validating the practical necessity of distributed computing infrastructure for modern data science applications.

The Rest-Mex sentiment analysis introduced the complexities of natural language processing in big data contexts, specifically addressing the dual challenge of massive text volumes and severe class imbalance in Spanish tourist reviews. The strategic combination of Ring Technique undersampling and Backtranslation augmentation, coupled with BETO fine-tuning, achieved 86.26\% accuracy on 208,051 reviews. This phase demonstrated that big data challenges extend beyond computational scalability to encompass sophisticated data preprocessing and balancing strategies, particularly crucial when working with naturally imbalanced real-world datasets.

The IMDb rating prediction represented the synthesis of all accumulated knowledge, integrating advanced feature engineering through intelligent contextual imputation across 85,855 films with 2,000 features each. The optimized XGBoost Regressor (RMSE=0.6001, R²=0.79) demonstrated that combining textual analysis via TF-IDF (5,000 features with HashingTF) with numerical features and sentiment analysis (46.83\% neutral, 32.65\% positive, 20.52\% negative) creates a robust predictive framework. The exploratory analysis revealed actionable insights: Christopher Nolan (8.22), Satyajit Ray (8.02), and Hayao Miyazaki (8.01) emerged as consistently top-rated directors, while Film-Noir (6.64), Biography (6.62), and History (6.54) represented the highest-rated genres—findings that illustrate how big data analytics can extract meaningful patterns from massive, heterogeneous datasets.

The comparative analysis between individual and distributed processing paradigms revealed a fundamental truth in big data analytics: the choice between local and distributed architectures is not merely about computational resources, but about the quality and completeness of the data available to the model. Linear models in distributed environments achieved higher predictive accuracy (95.04\% vs. 93.00\% AUC-ROC) precisely because they could leverage complete datasets, highlighting that in the big data era, data completeness can be as important as algorithm sophistication.

This chronological progression demonstrates that mastering big data requires a systematic approach: first establishing computational infrastructure and understanding algorithm scalability (Epsilon), then addressing domain-specific challenges like text processing and class imbalance (Rest-Mex), and finally synthesizing these capabilities to solve complex, real-world prediction problems (IMDb). Each phase introduced new technical challenges—from memory management and distributed computing, to natural language processing and intelligent imputation—building a comprehensive skill set essential for modern data science practitioners.

The practical implications extend beyond academic exercise. In an era where organizations generate petabytes of data daily, the ability to design scalable pipelines, implement distributed processing, handle imbalanced datasets, and extract actionable insights from heterogeneous data sources represents a critical competitive advantage.


\begin{thebibliography}{00}

\bibitem{chang2011libsvm}
C.-C. Chang and C.-J. Lin, ``LIBSVM: A library for support vector machines,'' 
\textit{ACM Transactions on Intelligent Systems and Technology}, vol. 2, no. 3, pp. 1--27, 2011.

\bibitem{zaharia2010spark}
M. Zaharia, M. Chowdhury, M. J. Franklin, S. Shenker, and I. Stoica, 
``Spark: Cluster computing with working sets,'' in 
\textit{Proc. 2nd USENIX Workshop on Hot Topics in Cloud Computing (HotCloud)}, 2010, pp. 10--10.

\bibitem{meng2016mllib}
X. Meng \textit{et al.}, ``MLlib: Machine learning in Apache Spark,'' 
\textit{Journal of Machine Learning Research}, vol. 17, no. 1, pp. 1235--1241, 2016.

\bibitem{breiman2001random}
L. Breiman, ``Random forests,'' 
\textit{Machine Learning}, vol. 45, no. 1, pp. 5--32, 2001.

\bibitem{cortes1995support}
C. Cortes and V. Vapnik, ``Support-vector networks,'' 
\textit{Machine Learning}, vol. 20, no. 3, pp. 273--297, 1995.

\bibitem{paszke2019pytorch}
A. Paszke \textit{et al.}, ``PyTorch: An imperative style, high-performance deep learning library,'' 
in \textit{Advances in Neural Information Processing Systems}, vol. 32, pp. 8026--8037, 2019.

\bibitem{kohavi1995study}
R. Kohavi, ``A study of cross-validation and bootstrap for accuracy estimation and model selection,'' 
in \textit{Proc. 14th Int. Joint Conf. Artificial Intelligence (IJCAI)}, vol. 14, no. 2, pp. 1137--1145, 1995.

\bibitem{goodfellow2016deep}
I. Goodfellow, Y. Bengio, and A. Courville, 
\textit{Deep Learning}. MIT Press, 2016.

\bibitem{verma2019}
H. Verma and G. Verma, 
``Prediction Model for Bollywood Movie Success: A Comparative Analysis of Supervised Machine Learning Algorithms,'' 
\textit{Review of Socionetwork Strategies}, vol. 13, pp. 155--172, 2019.

\bibitem{alvarez2023}
M. Á. Álvarez-Carmona \textit{et al.}, 
``Overview of Rest-Mex at IberLEF 2023: Research on Sentiment Analysis Task for Mexican Tourist Texts,'' 
\textit{Procesamiento del Lenguaje Natural}, vol. 71, pp. 425--436, 2023.

\bibitem{shalevshwartz2011}
S. Shalev-Shwartz, Y. Singer, N. Srebro, and A. Cotter, 
``Pegasos: Primal estimated sub-gradient solver for SVM,'' 
\textit{Mathematical Programming}, vol. 127, no. 1, pp. 3--30, 2011.

\end{thebibliography}

\end{document}